\documentstyle[psfig]{l-aa}

\def\ros{{\sl ROSAT }}
\def\asca{{\sl ASCA }}

\begin{document}
\thesaurus{03         
              (11.01.2;  
               11.09.1;  
               11.09.1;  
               11.09.2;  
               11.19.3;  
               13.25.2)  
}
   \title{\ros evidence for AGN and superwind activity in
NGC\,6240 and  NGC\,2782}
   \author{Hartmut Schulz \inst{1,2}, Stefanie Komossa  \inst{3},
Thomas W. Bergh{\"o}fer \inst{3,4} \and Berto Boer\inst{1}   
  }
\offprints{H. Schulz, Bochum address}
\institute{
Astronomisches Institut der Ruhr-Universit\"at, D-44780 Bochum, Germany
\and
Department of Physics and Astronomy, Dartmouth College,
Hanover, NH 03755, U.S.A.
\and
Max-Planck-Institut f\"ur Extraterrestrische Physik, Postfach 1603,
D-85740 Garching, Germany
\and
Space Sciences Laboratory, Berkeley, CA 94720, U.S.A.} 
\date{Received: 20 May 1997; accepted: October 1997}
   \maketitle
\markboth{Schulz et al.: NGC\,2782 and NGC\,6240}
{Schulz et al.: NGC\,2782 and NGC\,6240}

   \begin{abstract}
We present \ros observations of the starburst galaxy NGC\,2782
(HRI plus a weak PSPC frame) and the ultraluminous infrared
galaxy NGC\,6240 (PSPC). 
The (0.1--2.4) keV spectra of both objects appear similar.
However, due to better sampling spectral modeling is only
warranted in case of NGC\,6240 for which both a single thermal 
Raymond-Smith model ($kT = 0.44\pm0.15$) or a hybrid model
consisting of 80\% power-law with the canonical photon index $-1.9$
plus 20\% Raymond-Smith contribution ($kT = 0.63\pm0.35$) lead to
good fits. However, the single thermal model turns out to be
unlikely because it yields a luminosity of 
$3.8\,10^{43}$ erg s$^{-1}$, which is hard to reach in a 
starburst superwind--scenario. The hybrid model leads to a more
moderate luminosity of $5.2\,10^{42}$ erg s$^{-1}$, of which
$1.0\,10^{42}$ erg s$^{-1}$ can be attributed to shocked
superwind gas. We link the remaining $4.2\,10^{42}$ erg s$^{-1}$
powerlaw to an AGN component because the alternative of
inverse-compton scattering of the FIR radiation leads to a too low
flux when estimated with available data. The result appears to be
consistent with preliminarily announced ASCA observations.

For NGC\,2782 we find $L_x$(0.1--2.4 keV) = $4\,10^{41}$ erg s$^{-1}$
(within a factor of four) which can be explained by emission
from a shocked superwind region with a high preshock density
in agreement with earlier optical evidence for an outflowing supershell.  
  
\end{abstract}
%
\section{Introduction}
Seyfert (1943) included NGC\,2782 in his famous list of galaxies which
"show spectra having many high-excitation emission lines localized
in the nuclei".
However, Balzano's (1983) observations showed a more HII-region
like spectrum and, hence, she
classified NGC\,2782 as a starburst galaxy.
Nevertheless, early high-resolution spectroscopy revealed
line-width differences (Sakka et al. 1973;
Kennicut et al. 1989) that were attributed to an additional
high-excitation component. Boer et al. (1992) identified the
latter component as an expanding shell driven by a series of
supernova explosions in the starburst nucleus. 
Kinney et al. (1984)
measured an Einstein IPC-band (0.2--4 keV)
luminosity of 1.2\,10$^{41}$ or 8\,10$^{41}$ erg s$^{-1}$
for assumed power-law flux distribution indices of 
$\alpha = 0.0$ and $-0.6$, respectively. Boer et al. found
the observed X-ray luminosity too high as compared to other indicators
of the strength of the nuclear starburst. Therefore a
pointed \ros observation was proposed by B.B. and H.S., which
will be discussed in the present work.

NGC\,6240 is a morphologically peculiar and probably
interacting galaxy with two apparent nuclei (Fried \& Schulz 1983).
A more than 8 kpc extended central region reveals a
pronounced LINER-like spectrum and unusually
broad (FWHM 200--900 km s$^{-1}$) emission lines.
(Fosbury \& Wall 1979; Fried \& Schulz 1983;
Morris \& Ward 1988,
Keel 1990, Heckman et al.\ 1990, Veilleux et al.\ 1995,
Schmitt et al.\ 1996). 
After detection of its huge far-infrared 
luminosity of $\sim 10^{12} L_\odot$ (Wright et al. 1984) 
NGC\,6240 has been considered as a super starburst galaxy
(Joseph \& Wright 1985; Rieke et al.\ 1985).
In this vein, Heckman et al. (1987, 1990) suggested the presence
of superwind activity. Recently, ISO observations confirmed
the presence of a young starburst (Lutz et al.\ 1996).

However, there are suspicions of alternative or additional
power sources like an obscured AGN (de Poy et al.\ 1986) or
even ordinary starlight (Thronson et al.\ 1990). While the
latter hypothesis has been ruled out (Shier et al.\ 1996) there is
recent evidence from {\em Hubble Space Telescope} data 
for a high-excitation component indicative of
a hidden AGN or a young mini starburst (Barbieri et al.\ 1993,
Rafanelli et al.\ 1997).
Also, preliminary X-ray data from \asca suggest the presence
of an AGN (Mitsuda 1995).

To date, no further X-ray data have been published for NGC\,6240.
Rieke (1988) inspected HEAO-1 data and gave an upper limit
of $4.2\,10^{-12}$  erg s$^{-1}$ cm$^{-2}$ corresponding
to $L_x \le 1.0\,10^{43}$ erg s$^{-1}$ in the (2--10) keV range.
The \ros-PSPC data discussed below were taken from the archive.

Both objects have characteristics of a nuclear starburst and a
possible AGN in common. An AGN could be important for the
energy budget, in particular for the tremendous amount of
FIR radiation from an ultraluminous infrared galaxy (ULIRG). Perhaps
even more interesting, such an object might be the birthplace
of a newly forming AGN as is suggested by the similar
space densities of ULIRGs and QSOs (Sanders et al.\ 1988).

X-rays ought to be helpful in distinguishing between these two
types of nuclear sources. Sources powered by a starburst wind 
shocking the dense shell and
other ambient gas are usually expected to emit
a soft thermal X-ray spectrum with a (0.1--2.4 keV) luminosity 
$\la 10^{42}$ erg s$^{-1}$ while objects exceeding
this limit commonly show AGN tracers (e.g. Moran et al. 1994; Wisotzki \& 
Bade 1997).

A power-law X-ray spectrum with a photon spectral index $\Gamma = -1.9$
would be indicative of an AGN 
(e.g., Pounds et al.\ 1994, Nandra et al.\ 1997; 
Svensson 1994) (although AGNs may have thermal
components or starburst components added). The luminosity of a 
typical type-1 AGN in the \ros band is about 3\% of its bolometric 
luminosity.    

Luminosities given below are calculated via
plain application of the Hubble law with
$H_0 = 50$ km s$^{-1}$ Mpc$^{-1}$ yielding $d = 51$ Mpc
and 144 Mpc for NGC\,2782 and NGC\,6240, respectively
(when comparing luminosities or scales 
note that several recent papers on NGC\,6240 
assumed $d \approx 100$ Mpc which would halve the luminosities
given here).

\section{Observations and data reduction}
The data presented here were taken with the High Resolution Imager (HRI) 
and the
Position Sensitive Proportional Counter (PSPC) on board of the X-ray 
satellite \ros.
A detailed description of the \ros project, the 
satellite and its instrumentation can be found in Tr\"umper (1983), 
Tr\"umper 
et al. (1991), Pfeffermann et al. (1987) and Zombeck et al. (1990).
In brief, the \ros satellite observes in the soft X-ray range between
0.1 and 2.4 keV. The spatial resolution of frames obtained with the HRI 
and PSPC is about
$5\arcsec$ and $25\arcsec$, respectively. Unlike the HRI, the PSPC has 
sufficient
spectral resolution for obtaining information about the spectral shape. 
\subsection {NGC\,2782}
On April 18--20, 1992, a deep HRI X-ray image of NGC\,2782 was taken with 
a total effective 
exposure time of $\sim$ 21.7 ksec.
Via a standard source-detection procedure carried out 
with the EXSAS X-ray analysis software 
package (Zimmermann et al.\ 1994) we detected seven X-ray sources 
($>3\sigma$) in 
the field of view.  In Table 1 the detections are summarized. 
The positions, the background 
subtracted HRI count rates, and the optical identifications of the 
sources are given.
\begin{table*}[htbp]
  \caption{\label{ngc2782_tab} ROSAT HRI detections in the 
NGC\,2782 frame}
  \begin{flushleft}
  \begin{tabular}{lcccccccc}
  \hline
\noalign{\smallskip}
No. & \multicolumn{3}{c}{RA (2000)} & \multicolumn{3}{c}{Dec (2000)} & 
$10^{-3}$ cts/s
& optical identification\\
\hline
\hline
\noalign{\smallskip}
1 & $9^h$ & $14^m$ & 5\fs0&    +40\degr  &6\arcmin & 49\farcs5 & 
6.42$\pm$0.58 &
 NGC 2782 core (2\arcsec)\\
 2 & $9^h$ & $14^m$ & 4\fs0&   +40\degr  &7\arcmin & 40\farcs5 & 
1.39$\pm$0.29 & \\
 3 & $9^h$ & $14^m$ & 23\fs8&  +40\degr  &5\arcmin &  4\farcs9 & 
0.65$\pm$0.19 & gsc2987.00552
$V=13\fm5$ (18\arcsec)\\
 4 & $9^h$ & $13^m$ & 38\fs8 & +40\degr  &5\arcmin & 49\farcs3 & 
0.98$\pm$0.26 & \\
 5 & $9^h$ & $13^m$ & 52\fs4 & +40\degr &13\arcmin & 15\farcs9 & 
0.82$\pm$0.24 & \\
 6 & $9^h$ & $14^m$ & 26\fs5 & +40\degr  &0\arcmin & 23\farcs3 & 
0.75$\pm$0.23 & gsc2987.01040
$V=13\fm9$ (2\arcsec)\\
 7 & $9^h$ & $14^m$ & 46\fs3 & +40\degr  &3\arcmin & 6\farcs0 & 
12.29$\pm$0.79 & \\
      \hline
   \end{tabular}
   \end{flushleft}
\end{table*}
%
%

  \begin{figure}[thbp]
\vbox{\psfig{figure=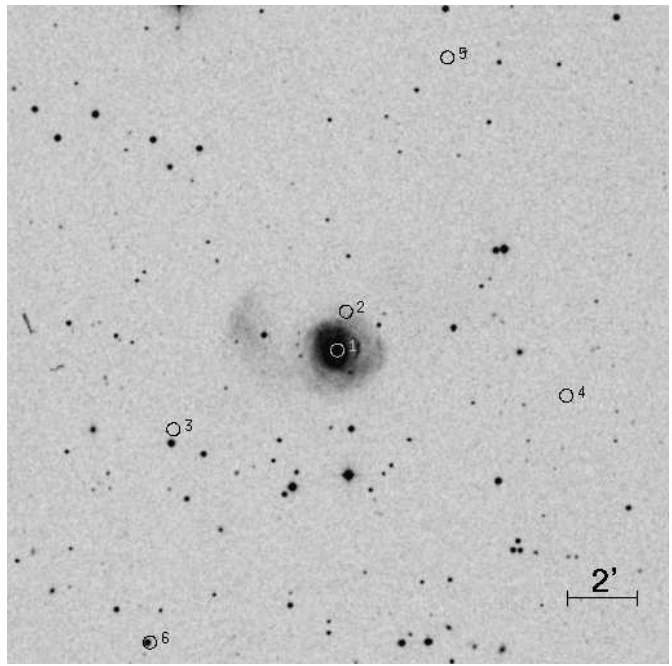,width=8.8cm,height=8.8cm,clip=}}\par
\caption{\label{ngc2782_ima} X-ray sources in the vicinity of NGC 2782 
plotted over the optical image}
\end{figure}

In Fig.\ 1 we plotted the X-ray source positions over the
optical image of NGC\,2782 taken from the digitized 
Palomar Sky Survey plates. The
source numbers are identified in Table 1.
The central source No.\ 1 is associated with the center of NGC\,2782. 
A four times fainter source (No.\ 2) 
appears 88\arcsec~north of the center of the galaxy.
The association of No.\ 2 with NGC\,2782 is unclear. It could be
a foreground or background object. The X-ray light curves of both sources
do not show any significant evidence for variability.

In addition to the HRI observation,  
NGC\,2782 appears on
an archived \ros PSPC frame (\ros observation number: WG701555P),
taken on Nov.\ 8, 1993. 
During this 7 ksec exposure NGC\,2782 was far off-axis
in the field of view. Due to the reduced spatial resolution of the 
telescope
away from the axis, sources No.\ 1 and 2 could not be separated. 
However, since No.\ 1 dominates 
the integrated X-ray flux crude information about its spectral shape
can be retrieved from the PSPC frame.

To this end we extracted the source photons within a 
radius of 300\arcsec, corrected the data for telescope vignetting and 
detector 
dead-time, and binned the PSPC spectrum according to a constant 
signal-to-noise. For the background correction of the source spectrum we 
selected the counts of a 'source free' region near NGC\,2782. 
We then fitted the background subtracted PSPC spectrum with a single 
powerlaw 
taking into account the Galactic interstellar absorption towards
NGC\,2782 of $N_{\rm H}^{\rm Gal} = 0.18 \times 10^{21}$ cm$^{-2}$ 
(Dickey \& Lockman 1990). The best fit yields a photon index 
$\Gamma = -3.1\pm1.1$ and an X-ray luminosity (0.1-2.4 keV) of 
log($L_x/{\rm erg\,s}^{-1}$)=41.6$\pm$0.6. This does not mean that the
spectrum actually {\em is} a powerlaw, but a spectral shape has to be 
assumed to obtain
a flux.  
\subsection {NGC\,6240} 
NGC\,6240 was observed twice with the \ros PSPC, 
on Sept. 2--4, 1992 and Feb. 13--14,
1993.   
The source was located at the center of the field of view. 
The total exposure time is about 8 ksec.

For the data analysis, source photons were
extracted within a circle of radius 150\arcsec.  
After removing all detected
sources within the inner field of view the background was determined 
and subtracted. The data were corrected for
vignetting and dead-time following standard prescriptions of the
EXSAS software (Zimmermann et al.\ 1994).
The source-countrate is about 0.06 cts s$^{-1}$.
For the spectral analysis source photons in the amplitude channels 11-240 
were binned
so that a constant signal/noise ratio of $5 \sigma$ was secured.
For the temporal analysis we chose time bins of 400 s which are the 
smallest ones
to smear out the variations caused by the satellite's wobble motion. 

The X-ray isophotes for the 1992 exposure overlayed on the digitized POSS 
are
shown in Fig.\ 6.   
The K5 star HD\,152140 is the strongest X-ray source in both observations 
(mean countrate $\sim 0.3$ cts s$^{-1}$).
All other sources in the field are significantly weaker than NGC\,6240.   
The K star was used to positionally align both frames. 

\section{Spatial analysis for the NGC\,6240 frames}
Approximating the source photon distribution by a Gaussian
and representing the extent by the FWHM of this distribution,
we find a value of 37\arcsec.  
This exceeds the width of the PSPC point spread functions for
different channels. E.g., for a point source we
determined FWHMs of 23\arcsec~ and 27\arcsec~ 
at channel 110 and 60, respectively.
A notable extension beyond the PSF is present
in all energy intervalls (soft as well as hard).
To be sure, we additionally verified that the extent is not due to the
electronic ghost imaging that widens the point spread function below 0.2 
keV
(Nousek \& Lesser 1993).
Consequently, the X-ray image obtained with the PSPC is extended
in both observations.

The X-ray positions of both, NGC\,6240 and the K star, agree with 
their optical positions (as given in SIMBAD) within less than one arcsec
for the 1993 observation. In the 1992 data, {\em both} X-ray positions
are shifted by $\sim$ 10\arcsec~ relative to their optical positions.   
This shift can be traced back to the known boresight error 
of the telescope (Briel et al.\ 1994) and has been corrected for in the 
overlay of the X-ray contours on the optical image (Fig.\ 6).

\section {Temporal analysis for the NGC\,6240 frames}
Figure 2 shows the individual countrates versus time for the
1992 and 1993 observations. Given the statistical errors no
significant variations can be claimed. In the \ros all-sky survey,
NGC\,6240 was detected with a countrate of $0.086\pm0.016$
(Voges et al.\ 1997) which is consistent with the individual
pointed observations.
%
%
%
  \begin{figure}[thbp]
      \vbox{\psfig{figure=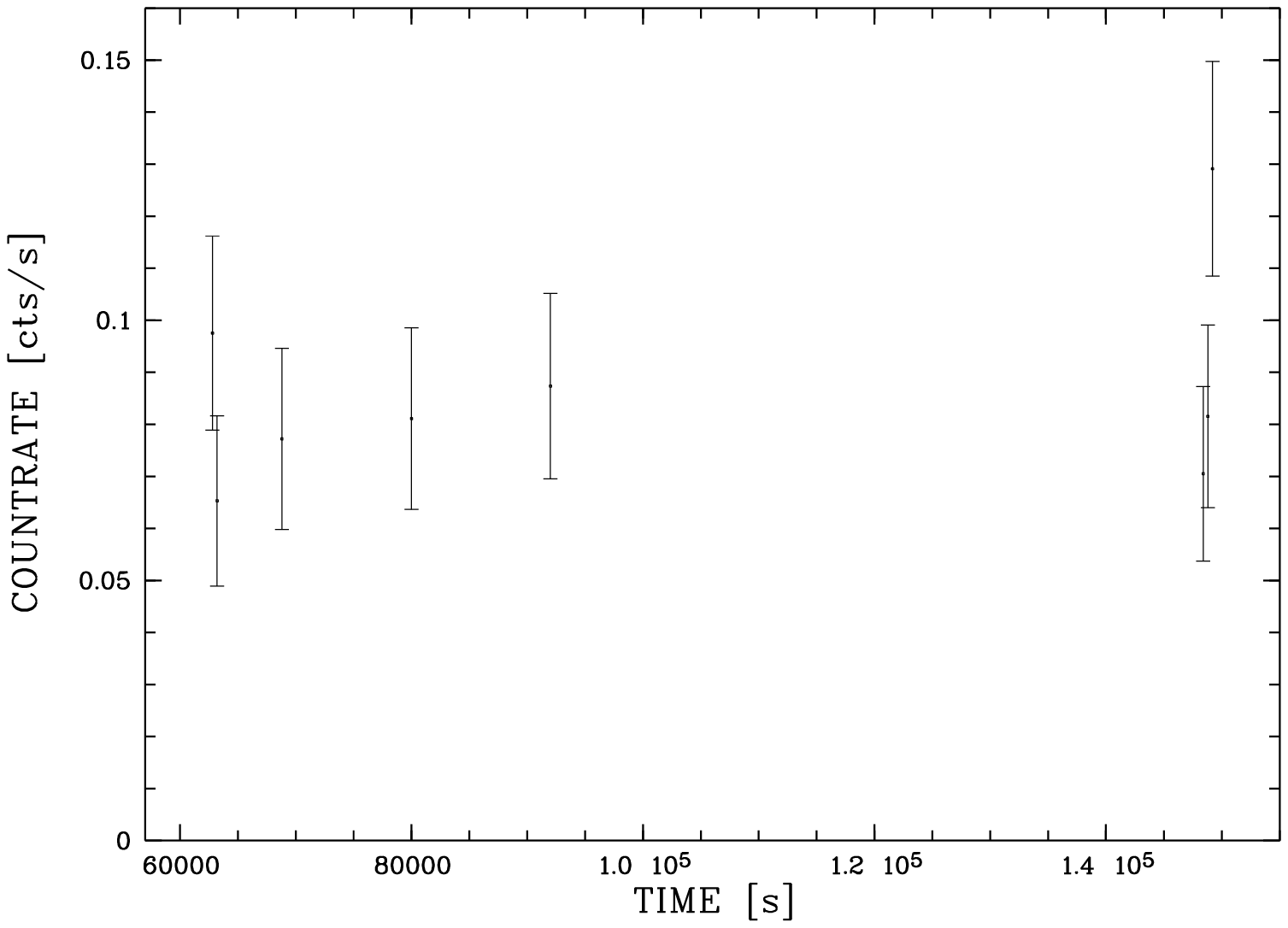,width=8.5cm,height=3.8cm,%
          bbllx=2.9cm,bblly=1.1cm,bburx=18.1cm,bbury=12.2cm,clip=}}\par
      \vspace{0.3cm}
     \vbox{\psfig{figure=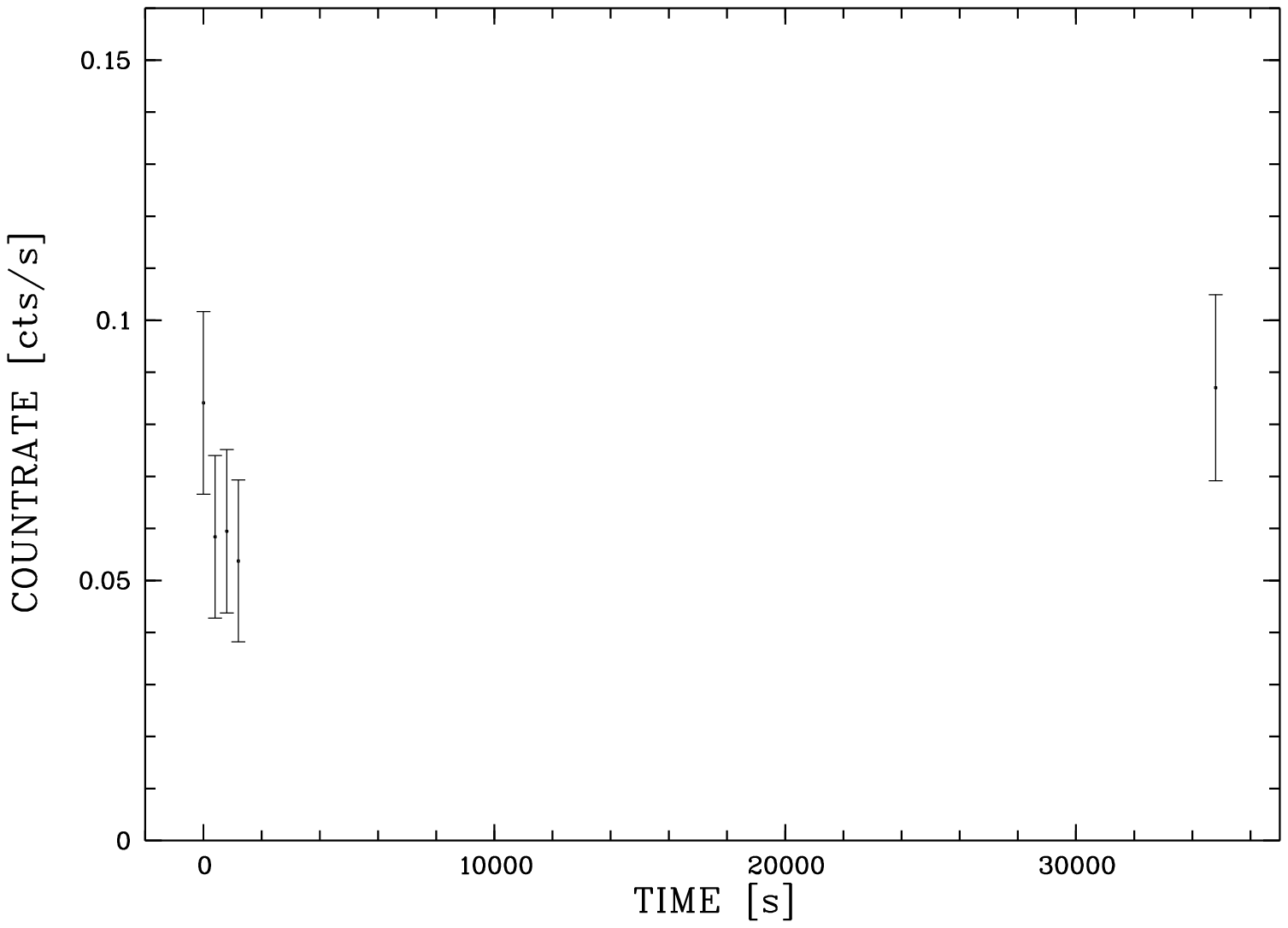,width=8.5cm,height=3.8cm,%
          bbllx=2.9cm,bblly=1.1cm,bburx=18.1cm,bbury=12.2cm,clip=}}\par
\caption[light]{X-ray lightcurve of NGC\,6240. 
The time is measured in seconds from the start of the observations. 
Upper panel: 1992 observation,
lower panel: 1993. The bin size in time is 400 s.}
\label{light}
\end{figure}
\section{Spectral analysis for NGC\,6240}
As shown in Fig.\ 3 there is
no evidence for spectral changes between the 1992 and
1993 observations so that the data were merged
to ensure better statistics.
%
%
  \begin{figure}[thbp]
    \vbox{\psfig{figure=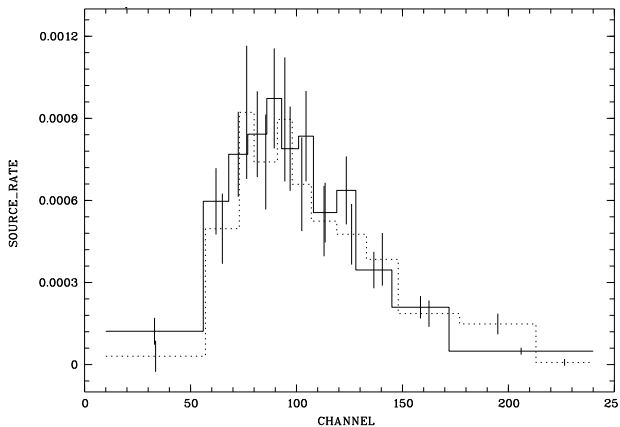,width=8.2cm,
        bbllx=2.4cm,bblly=1.1cm,bburx=18.1cm,bbury=12.09cm,clip=}}\par
\caption[vgl]{ Comparison of the spectra of both observations
(solid line: 1992 obs., dotted: 1993), countrate [ph/s/bin] vs. detector
channel number 
spanning the range (0.1 -- 2.4) keV. The spectra are
the same within the error bars.
}
\label{vgl}
\end{figure}

However, the $\sim400$ photons detected in total limit any 
meaningful fits by spectral models to no more than two spectral 
components.
Even with this restriction, the measured total soft X-ray luminosity 
is highly model dependent as is illustrated by the numerous
statistically acceptable fits in Table 2.
This is due to the large bin widths which causes individual fitted
photon numbers in each bin to depend on the spectral shape.

In addition to the individual model determining parameters like 
temperature
$T$ (one parameter of the ``thermal'' models as bb = black body, thb = 
thermal
bremsstrahlung, rs = thin thermal plasma in collisional equilibrium as
computed by Raymond \& Smith (1977)) or powerlaw index $\Gamma$
(for ``nonthermal'' pl = powerlaw models according to a photon flux
distribution $\Phi_{\rm photon} \propto E^{\Gamma}$ with $E$ = photon 
energy)
and the necessary normalization constants, the fits depend on the
column density $N_{\rm H}$ of ``cold'' gas that is represented by the 
summed
photoelectric cross sections for gas with cosmic chemical abundances.
Table 2 shows that individually fitted $N_{\rm H}$ values can grossly vary
among the fitted models because $N_{\rm H}$ influences the 
slope of the soft component. 

We adopt the Galactic hydrogen column $N_{\rm Gal} = 0.549\,10^{21}$ 
cm$^{-2}$
(Dickey \& Lockman 1990) in the
direction of NGC\,6240 as a lower limit. However, a spectral emission
model fitted with this column would have to encounter no absorption
{\em within} NGC\,6240 which appears somehow unlikely in view of the
abundant molecular and ionized material observed in the central region 
of this object. The emission would have to arise in an ``outer shell''
or penetrate through optically thin windows in the ISM. Indeed,
a pl-model with $N_{\rm H} = N_{\rm Gal}$ (model E in Table 2) leads to a 
bad fit while
a bb-model (model A) fits well.
However, such a huge bb-shell, optically thick in soft X-rays and
extending over more than 30\arcsec~or, in linear scales, over more than
21 kpc cannot be reasonably imagined in physical terms. The physically 
expected 
thin X-ray emitting gas will emit lines with a significant contribution
so that the thb fit (model B) is excluded as well. 
Our main goal to showing these (from the beginning) unrealistic models is 
to find out
the range of luminosities (last column in Table 2) obtained by different 
spectral shapes.

A powerlaw with free column (model D) is statistically feasible but has a 
steep
slope with $\Gamma = -3.7$. 
Only the subgroup of so-called narrow-line Seyfert-1 galaxies (NLSy1; e.g.
Puchnarewicz et al.\ 1992; Boller et al.\ 1996;
Greiner et al.\ 1996) is known to have comparatively steep slopes.
Komossa \& Greiner (1995) proposed to
explain these objects by warm absorption of an intrinsically flat
($\Gamma_x \simeq -1.9$) powerlaw. The steep spectrum is then mimicked 
by the presence
of matter-bounded, completely H-ionized absorbing gas (so-called ``warm 
absorbers''),
which absorb by metal ions. 
Applying a standard warm-absorber model (as e.g.\ described in Komossa \& 
Fink 1997) 
we found no acceptable fit. 

Hence, the steep slope, far off the
canonical value $\Gamma = -1.9$, 
has to be intrinsic which, if persisting in the UV,
would cause an ultraluminous UV component never seen in any unobscured
object. Finally, the luminosity of model D (last column in Table 2) of
$\sim 10^{44}$ erg s$^{-1}$ is so huge that, for common AGN $L_x/L_{\rm 
bol}$
ratios, a hidden QSO of $> 10^{46}$ erg s$^{-1}$ would be required
that would heat the central dust in NGC\,6240 to more than three times the
observed FIR luminosity unless the bulk of its radiation escapes 
elsewhere.
We consider model D as unlikely.
     
We are left with model C as a purely thermal solution
and F as a hybrid model consisting of thin thermal gas plus a nonthermal
component. The added pl in F leads to a comfortably small luminosity
compared to C because its assumed `flat' slope helped to minimize
$N_{\rm H}$. Tuning the pl index to flatter slopes
does not lead to a significant further luminosity reduction (cf. model E).
Possible origins of the components will be discussed in Sect.\ 6.

%
  \begin{figure}[thbp]
      \vbox{\psfig{figure=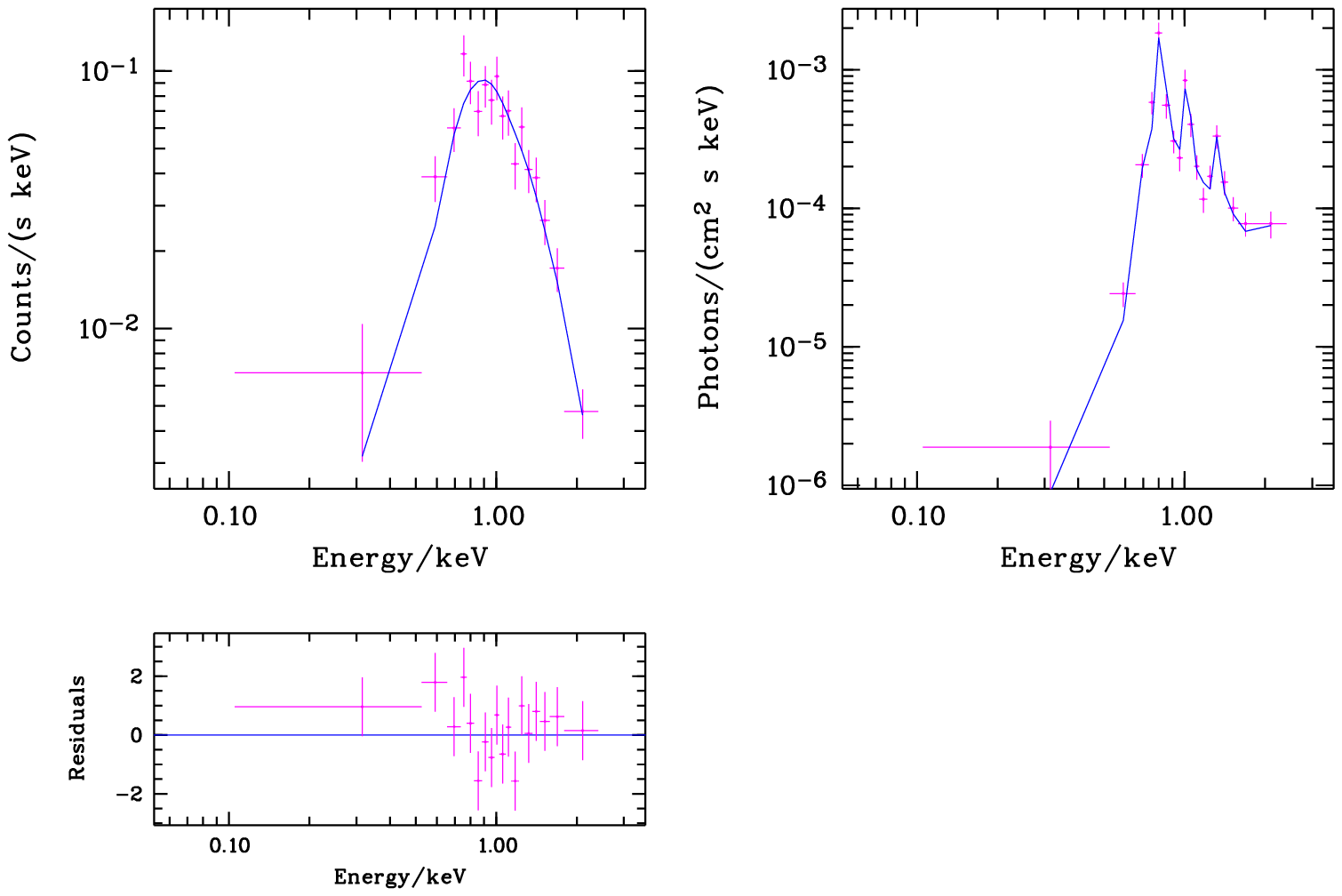,width=8.0cm,%
          bbllx=2.4cm,bblly=1.1cm,bburx=10.1cm,bbury=11.6cm,clip=}}\par
\caption[SEDx]{The upper panel shows the observed X-ray spectrum of NGC 
6240
(crosses) and the best-fit Raymond-Smith model (solid line).
The residuals of this fit are displayed in the next panel.
}
\label{SEDx1}
\end{figure}
  \begin{figure}[thbp]
      \vbox{\psfig{figure=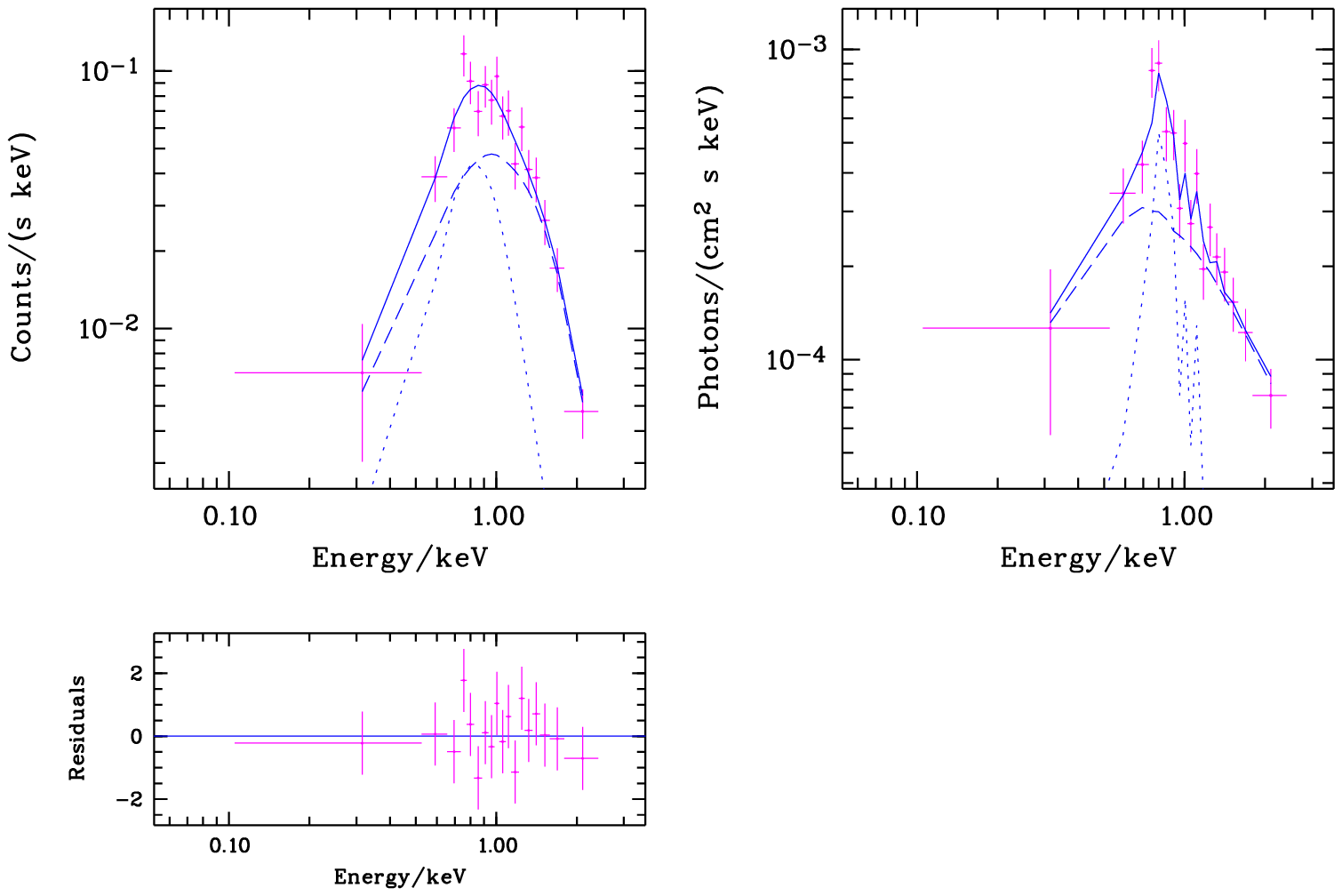,width=8.0cm,%
          bbllx=2.4cm,bblly=1.1cm,bburx=10.1cm,bbury=11.6cm,clip=}}\par
\caption[SEDx]{The upper panel shows the observed X-ray spectrum of NGC 
6240
(crosses) and the best-fit rs+pl model (solid line). 
The residuals of this fit are given in the lower panel. 
}
\label{SEDx2}
\end{figure}

   \begin{table*}             
     \caption{Spectral fits to NGC\,6240 (pl = power law, bb = black 
body, thb =
thermal bremsstrahlung, rs = Raymond-Smith model of cosmic abundances).
 $f_{\rm (0.1-2.4) keV}$ = 0.1 -- 2.4 keV flux corrected for
absorption, $T$ = temperature
 of bb, thb or rs component, $N_{\rm H}$ = cold absorbing column,
d.o.f. = degrees of freedom.
Instead of individual error bars, we list different models that
successfully describe the data.
              }
     \label{fitres}
      \begin{tabular}{lllcllccc}
      \hline
      \noalign{\smallskip}
        model & $N_{\rm H}$ & $\Gamma_{\rm x}$ & Norm$_{\rm pl}$ & $T$
                            & Norm$_{\rm other}$ & $f_{\rm (0.1-2.4) keV}$
                            & $\chi^2_{\rm red}(d.o.f)$
                            & $L_{\rm (0.1-2.4) keV}$ \\
       \noalign{\smallskip}
      \hline
       \noalign{\smallskip}
              & 10$^{21}$ cm$^{-2}$ & & ph/cm$^2$/s/keV$^{(1)}$ & keV & 
$^{(2)}$
& erg/cm$^2$/s &  & erg/s\\
       \noalign{\smallskip}
      \hline
      \hline
      \noalign{\smallskip}
A (bb) & 0.549$^{(3)}$ & -- & -- & 0.28 & $8.59\,10^{-4}$ & 
$9.97\,10^{-13}$ & 0.98 (16) & $2.5\,10^{42}$ \\
B (thb) & 1.96 & -- & -- & 0.73 & $6.73\,10^{-4}$ & $3.10\,10^{-12}$ & 
0.89 (15)& $7.7\,10^{42}$\\
C (rs)  & 9.63 & -- & -- & 0.44 &  &$1.52\,10^{-11}$ & 1.11 (15)& 
$3.8\,10^{43}$\\
\hline
\noalign{\smallskip}
D (pl) & 3.5 & $-3.7$ & $1.00\,10^{-3}$ & -- & -- & $4.42\,10^{-11}$ & 
0.86 (15)& $1.1\,10^{44}$\\
E (pl) & 0.549$^{(3)}$ & $-1.5$ & $3.65\,10^{-4}$ & -- & -- & 
$1.44\,10^{-12}$ & 2.55 (16) & $3.6\,10^{42}$\\
F (rs+pl) & 1.61 & $-1.9^{(4)}$ & $3.63\,10^{-4}$ & 0.63 &  & 
$2.09\,10^{-12}$ & 0.77 (14)&$5.2\,10^{42}$\\
      \noalign{\smallskip}
      \hline
      \noalign{\smallskip}
  \end{tabular}

\noindent{\small $^{(1)}$ at 1 keV;  $^{(2)}$ ph/cm$^2$/s for bb,
ph/cm$^2$/s/keV at 1 keV for thb;  
$^{(3)}$ fixed to the Galactic value; $^{(4)}$ fixed \\
}
  \vspace{-0.4cm}
   \end{table*}

\begin{figure*} 
%
\vspace{0.4cm} 
\vbox{\psfig{figure=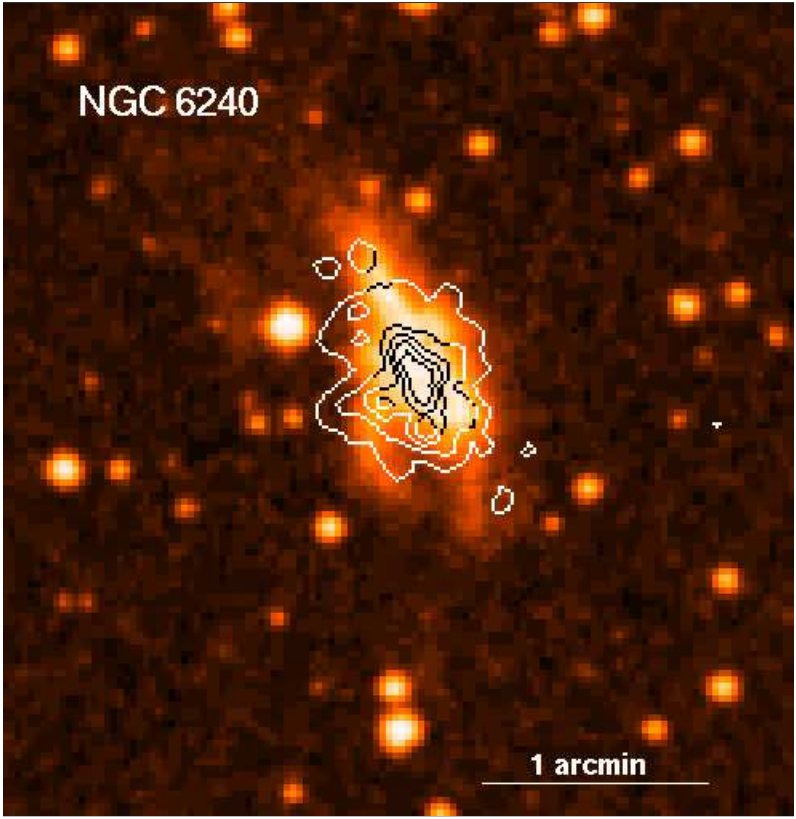,width=10.5cm,
bbllx=3.15cm,bblly=10.45cm,bburx=18.1cm,bbury=26.1cm,clip=}}\par
\hfill
\begin{minipage}[]{0.35\hsize}\vspace{-3.0cm}
\hfill
\caption[over]{Overlay of the X-ray contours   
on the optical image of NGC\,6240 from the digitized Palomar sky survey.  
}
\label{over}
\end{minipage}
\end{figure*}
\section{Discussion}
Due its more complete database we first discuss NGC\,6240.
\subsection{NGC\,6240}
From the discussion in Sect.\ 5 only model C (thermal rs with
luminosity $3.8\,10^{43}$ erg s$^{-1}$) and
the pl+rs model F (rs luminosity $1\,10^{42}$ erg s$^{-1}$, pl luminosity
$4.2\,10^{42}$ erg s$^{-1}$) of Table 2 are acceptable. 
Given the abundant evidence for superwind activity and a possible
AGN these appear to be the most likely sources for the X-ray emission.
A notable contribution by
direct emission from OB stars present in the supposed starburst is ruled 
out:
each star only emits $10^{32-33}$ erg s$^{-1}$
of a rather soft X-ray energy distribution 
(e.g. Vaiana et al.\ 1981, Bergh{\"o}fer et al.\ 1997). A 
significant fraction of $10^{10-11}$ unobscured O stars would have shown
up in the optical. 

A large number of massive X-ray binaries and a few supernova remnants 
with luminosities in the range 
10$^{35-38}$ erg s$^{-1}$ do collectively play a significant role in 
`normal' spiral
galaxies. NGC\,6240 might be the product of the collision of such objects
which in their normal state have a X-ray luminosity of $10^{39-40}$ erg 
s$^{-1}$
(Fabbiano 1989).
Since there is no known mechanism to boost the number
of X-ray binaries by two orders of magnitudes in the interaction process
it appears practically ruled out that the  
$10^{42-43}$ erg s$^{-1}$ soft X-rays come from binaries.
Supernova remnant activity can indeed be boosted in a starburst triggered 
by
the interaction, but this leads us directly to the superwind model.
\subsubsection{Superwind estimate}
Sophisticated two-dimensional hydrodynamical simulations of galactic
superwinds are available (e.g.\ Suchkov et al.\ 1994). However, such 
studies
mainly address the ``prototype'' M\,82 so that their particular examples
hardly reach the temperatures and luminosities found in our fits.
Lacking precise constraints our goal is limited to the estimate 
whether the huge thermal rs-component
luminosities from the fits are feasible. Hence, we use a scalable
theory rather than a hydrodynamical model with specific initial 
conditions.
We utilize the most sophisticated models of this type which were built by
Mac Low \& McCray (1988; for an application see also Schulz 1990)
who had extended the seminal work by Weaver et al.\ (1977).
 
Densities and distribution functions of the ISM in the center of 
NGC\,6240 are
not known. The extent of the supershell can be estimated from the 
five-finger 
structure on available
H$\alpha$ images (Schulz 1984, Keel 1990, Thronson et al.\ 1990) that 
yield
$\sim10$ kpc including a correction for a disk inclined by up to 40\degr~ 
from edge-on. 
The strong dust lane on B photographs (best visible in the masked prints 
made 
by Fosbury \& Wall 1979) 
suggests a close-to-edge-on view of the main disk.

The large line widths (FWHM) cited in Sect.\ 1 which most frequently lie 
in the range
between 480 and
620 km s$^{-1}$, but occasionally reach 900 km s$^{-1}$~(Keel 1990), 
are due to the summed effect of 
outflow, rotation and `turbulence' (the filamentary shell in M\,82 shows
linewidths of up to 300 km s$^{-1}$; cf. Bland \& Tully 1988).
As pointed out by van der Werf et al.\ (1993),
shock velocities exceeding 200 km s$^{-1}$ would boost the [OI] and [SII] 
emission-line
intensities relative to H$\alpha$ and lower [OIII]/[OI] thereby 
deteriorating the explanation of
the optical spectra by a large contribution of shockheated gas. 
Since $\sigma = {\rm FWHM}/2.35$, an outflow velocity
of the shell of 200 km s$^{-1}$ is consistent with the data and turbulence
components with $\sigma \sim 150$ km s$^{-1}$ (FWHM $\sim 350$ km 
s$^{-1}$).  

As mean density of the swept up ISM we have chosen
$n_0 = 0.1$ cm$^{-3}$, higher than the canonical values of $10^{-2}$ 
cm$^{-3}$
from our Galaxy or M\,82 (Strickland et al.\ 1997). The most prominent 
extended bright
parts of the supershell are on the eastern side of the double nucleus 
which might
be due to a slightly asymmetric location of the sources of the wind.

The mechanical input power $L_{\rm mech}$ is given by the supernova rate 
(MacLow \& McCray 1988)
for which we adopt 1 SN yr$^{-1}$ (van der Werf et al.\ 1993 estimate 2 
yr$^{-1}$;
Colbert et al.\ 1994 obtain 1 yr$^{-1}$ from the diffuse radio emission 
but are
uncertain about the extra power released in the compact radio knots). 
With an energy
per SN of 10$^{51}$ erg, we obtain
$L_{\rm mech} = 3\,10^{43}$ erg s$^{-1}$. According to Heckman et al.\ 
(1996) shock models
by Binette et al.\ (1985) predict $L$(H$\alpha) \sim 10^{-2} L_{\rm 
mech}$, yielding
$L_{\rm mech} \sim 10^{44}$ erg s$^{-1}$ with $L$(H$\alpha) = 
7.8\,10^{41}$ erg s$^{-1}$ from
Fosbury \& Wall (1979). Largely due to a larger reddening correction, 
Veilleux et al.\ (1995) obtain $L$(H$\alpha) = 4.3\,10^{42}$ erg s$^{-1}$ 
which would lead to
$L_{\rm mech} \sim 4\,10^{44}$ erg s$^{-1}$ or 14 supernovae yr$^{-1}$ 
which is
not compatible with other evidence. 
We take 3 SN yr$^{-1}$ or $L_{\rm mech} \sim 10^{44}$ erg s$^{-1}$ as an 
upper limit
and assume that H$\alpha$ has to be partially attributed to ionizing 
sources other
than shocks.
 
Scaling of the formula of MacLow \& McCray (1988) leads to the following
expressions (for $L_x$ we followed Heckman et al.(1996) who had carried 
out
the integration over the superbubble model in the \ros band):
\begin{equation}
R = (10.2\, \mbox{\rm kpc})\, L_{3E43}^{1/5}\, 
n_{0.1}^{-1/5}\, t_{3E7}^{3/5}
\end{equation}
\begin{equation}
\frac{{\rm d}R}{{\rm d}t} = (200\,\mbox{\rm km s}^{-1}) 
\,L_{3E43}^{1/5}\, 
n_{0.1}^{-1/5}\, t_{3E7}^{-2/5}
\end{equation}
\begin{equation}
L_x = (8.8\,10^{41} \mbox{\rm erg s}^{-1})\, 
L_{3E43}^{33/35}\, 
n_{0.1}^{17/35}\,t_{3E7}^{19/35}
\end{equation}
where $n_{0.1} = n_0/(0.1 \mbox{\rm cm}^{-3})$,
$t_{3E7} = t/(3\,10^7 \mbox{\rm yr})$
and $L_{3E43} = L_{\rm mech}/(3\,10^{43} \mbox{\rm erg s}^{-1})$.

These relations show that, with the observational constraints so far
available, it is quite natural to account for the thermal $L_x \sim 
10^{42}$
erg s$^{-1}$ obtained for the rs-component of the
rs+pl hybrid model.

However, how can we reach the $3.8\,10^{43}$ erg s$^{-1}$
required for the single-rs fit? We need a factor of 43 in eq.\ 3.
A factor-of-3 increase of $L_{3E43}$ would fit to our upper 
limit of $L_{\rm mech}$ that is in accordance with
shocked H$\alpha$ (see above) but already strain the conditions for
$R$ and $\frac{{\rm d}R}{{\rm d}t}$ (Eqs.\ 1 and 2). Enhancing the
density to $n_{0.1} = 10$ (a rather high value in the halo into which
the bubble expands) would boost $L_x$ by a factor 3.5 and comfortably 
decrease
$R$ and $\frac{{\rm d}R}{{\rm d}t}$ to still acceptable values. 
However, we still need another factor 4.3 requiring
a factor of 20 in the expansion-time parameter. However, a $6\,10^8$ yr
old bubble would have already commenced other starbursts for which
no trace is visible and, more seriously, our theory would no more be
applicable.

This exercise of boosting the parameters to high but not excludable
values demonstrates the basic difficulty of getting a soft-X ray
luminosity of several $10^{43}$ erg s$^{-1}$ from a single superbubble.

Some concern about the single-rs fit (model C in Table 2) also arises by 
the   
high cold absorbing column, which corresponds to $A_V = 5.4$ mag (using
Bohlin's et al.(1978) relation). Thronson et al. (1990) find $A_V \approx 
(3^m - 5^m)$ 
in the double nuclei but elsewhere $A_V \approx 1^m$. The latter value 
should
be more typical for most of the volume occupied by the extended 
supershell. 
\subsubsection{AGN contribution}
Inverse-Compton (IC) upscattered photons from the FIR photons in the 
radio sources of NGC\,6240 would generate a flux of
$\sim 10^{-14}$ erg s$^{-1}$ cm$^{-2}$ (Colbert et al.\ 1994). 
The powerlaw component of model F (Table 2) emits a flux of 
$1.69\,10^{-12}$ erg s$^{-1}$ and is therefore likely to be
attributed to an AGN-like source.

It may, however, appear striking that the expected IC scattered flux fits
our measured powerlaw
if the Compton X-ray emission disk-halo model devised 
by Goldshmidt \& Rephaeli (1995, GR95) is upscaled
from the starburst galaxy NGC\,253 to NGC\,6240 via the FIR ratio. 
The GR95 model predicts
an unattenuated Compton flux of $9.2\,10^{-15}$ erg cm$^{-2}$
s$^{-1}$ in the \ros band if NGC\,253 were placed at 144 Mpc, the distance
of NGC\,6240 adopted here. Scaling this by the ratio 175 of the FIR
luminosities (the flux formula from  Helou et al.\ (1985) yields
$9.75\,10^{11} L_{\odot}$ for NGC\,6240, GR95 use $5.58\,10^{10} 
L_{\odot}$
for NGC\,253) leads to $1.61\,10^{-12}$ erg cm$^{-2}$ s$^{-1}$ which
is, indeed, nearly the same as the power-law flux of NGC\,6240
given in the last paragraph.

However, this simple scaling may be misleading because the magnetic fields
in the radio sources of NGC\,6240 are 
higher than in NGC\,253 (Colbert et al.\ 1994). 
In the flux-dominating
two radio nuclei the {\em lower limit} of the magnetic field is fifty 
times ($> 4.4\,10^{-4}$ G) larger than that in the central synchrotron 
source 
of NGC\,253 ($9\,10^{-6}$G) while in the western extended regions
$B=2\,10^{-5}$ G which is 2.2 times larger. Granted that other conditions 
are scalable, the
boost in the magnetic energy densities ($\epsilon_{\rm mag} \propto B^2$)
should diminish the above NGC\,253-scaled IC flux estimate 
(roughly $\propto \epsilon_{\rm mag}^{-1}$) 
by factors between $<4.2\,10^{-4}$ (compact components) and $0.2$ 
(extended regions).
Hence, within this scheme it appears that at least the extended radio 
sources 
may contribute a significant comptonized power law. 

Adopting this scaled GR95 model for NGC\,6240 would, however, mean 
to scale the radiation energy density 
distribution from NGC\,253 by luminosity which requires an identical 
size of the FIR emitting region in objects that differ in luminosity
by a factor of hundred. Even if this might be the case 
the electron distribution in the extended halo components 
of NGC\,253 and NGC\,6240 would have to be similar as well, which is
doubtful because of the hundred times smaller propagation lengths for the
disk-halo connection via electron convection and 
diffusion in the latter galaxy. This is due to the increased radiative 
loss in the central disk (see, e.g., Eq.\ 8 and the subsequent discussion
in GR95).

In their more direct approach, Colbert et al.\ related 
the Compton upscattering to the {\em observed} synchrotron sources
and tested two complementary types of FIR distributions
(distributed like the radio sources and uniform distribution).  
Neither case, regardless of compact or extended
sources, yielded any flux above a few percent of that of our pl component.

Finding no direct clue from available observations 
that IC upscattering of FIR photons plays a significant role 
we continue to interpret the powerlaw by an AGN-like source.
Notwithstanding, GR95-type IC models for extended components might become 
relevant 
if some of the presently only crudely known parameters 
will attain more favorable
values for such processes in the future.          

The supershell model in Sect.\ 6.1.1 easily accounts for the rs-component
of model F which makes 20\% of the total flux. We now have to explain
the remaining $4.2\,10^{42}$ erg s$^{-1}$ in the powerlaw. Aside from the 
X-rays,
the only other hint for high-energy radiation of an AGN is given by the 
high-excitation
core in the southern nucleus (Rafanelli et al.\ 1997) that accounts for 
$\sim 8$\%
of H$\beta$. This component is probably due to a `window' in the ISM
and does not allow a direct view of the continuum source.

Consequently, the supposed AGN is likely to be seen in scattered light. 
This
is in agreement with our non-detection of variability of the supposed AGN
because scattering the X-rays over several kpc would smear 
out any short-term variability.
The electrons in the extended ionized regions
of the superwind and supershell itself are one possible source of 
scatterers. 
A simple geometry with only one constant-density 
population of electrons 
defines a fraction $f_s$ of the AGN luminosity $L_x$
to be seen as scattered luminosity $L_s = f_s  L_x$ by the condition
\begin{equation}
L_s = f L_x n_e \sigma R = f_s  L_x
\end{equation}
where $\sigma$ is the Thomson cross-section $0.665\,10^{-24}$ cm$^2$,
$L_s$ our measured powerlaw luminosity of $4.2\,10^{42}$ erg s$^{-1}$.
From the central biconical H$\alpha$ structure one estimates that
a fraction $f \sim 0.3$ can escape from the central $L_x$ (for simplicity 
assumed
to be isotropic in the first instance), corresponding to an ideal bicone 
with
opening angle $\sim 90^{\circ}$ for each side. In practice, the structure 
has a broader
base and has a shape something in between a biconical and a bycilindrical 
configuration,
but for a crude estimate this distinction is unimportant.
  
Basically two types of
scatterers are available: Firstly, the dense material close to the 
baseline and in the
broken shells. For isothermal shocks we would expect densities of the 
order of
$\sim 10^2$ cm$^{-3}$. For the bright filamentary material within the 
innermost
2 kpc an electron density of 500 cm$^{-3}$ was measured by
Heckman et al.\ (1990). Secondly, low-density, distributed in the bicone, 
could reach
significant electron-scattering depths due to its spatial extent of up to 
10 kpc.

The H$\alpha$ image from Keel (1990) shows a strong surface-brightness
enhancement in the southern
nucleus suggesting that 
luminous high-density material is concentrated there.
Adopting from the above $n_e = (2-5)\,10^2$ cm$^{-3}$ and $R = 200$ pc 
leads to
$f_s = (2.4 - 6.1)\,10^{-2}$ or a predicted (isotropic)
$L_x = (6.8\,10^{43} - 1.7\,10^{44})$ erg s$^{-1}$. The mass of this 
ionized  gas in
the inner bicone would be $\sim 10^{8} M_{\sun}$, compatible with 
$M_{{\rm H}^+} > 1.1\,10^{10} M_{\sun}/n_e$ estimated from hydrogen-line
luminosities (DePoy et al.\ 1986; Veilleux et al.\ 1995; the lower limit 
applies
for pure recombination). The amount of dense
ionized gas is only minute compared to the
$3.7\,10^{10} M_{\sun}$ of available cold molecular gas in the center of
NGC\,6240 (Solomon et al.\ 1997). Its filling factor is likely 
to be $ < 1$, but the covering fraction must be large to ensure an
{\em effective} scattering length of 200 pc. 
In this scenario the power-law component should be unresolved
with current X-ray telescopes.

The second possibility -- 
an extended low-density region -- would be much less efficient in 
scattering and 
involve a huge mass with unclear origin. 
E.g. $R = 5$ kpc, $n = 2$ cm$^{-3}$ would lead to 
$L_x = 6.9\,10^{44}$ erg s$^{-1}$ and $M_{{\rm H}^+} = 1.2\,10^{10} 
M_{\sun}$; 
in case of $R = 10$ kpc, $n = 0.1$ cm$^{-3}$, we would need 
$M_{{\rm H}^+} = 4.9\,10^{9} M_{\sun}$, but
$L_x = 6.8\,10^{45}$ erg s$^{-1}$ that would exceed $L_{\rm FIR}$ of 
NGC\,6240! 

We conclude that scattering is feasible but,
even in a simple high-density scenario, it requires
a rather luminous AGN in the range $L_x \sim 10^{43-44}$ erg s$^{-1}$, 
which
corresponds to $L_{\rm bol}({\rm AGN}) \sim 10^{44-45}$ erg s$^{-1}$ 
for a typical AGN continuum,
which is an appreciable fraction or nearly all of
the FIR luminosity ($3.7\,10^{45}$ erg s$^{-1}$) of NGC\,6240.

The luminosity demand for the AGN is lowered in case of
anisotropy of the AGN radiation because in the 
unified scenario more 
luminosity is expected along the bicone than in the equatorial plane of 
the molecular torus.
An alternative possibility would be
a component like a `warm scatterer', i.e.\ near-central high density gas 
that scatters
the X-rays into an escape window of the ISM but shows itself relatively
little intrinsic emission, is possible.

The need for an AGN in NGC\,6240 is also supported by preliminarily
reported \asca data (Mitsuda 1995) that show a power law in the 2--10 keV 
range
plus a strong emission-line complex around FeK$\alpha$.
\subsection{NGC 2782}
In Sect.\ 2.1 we obtained $L_x = 4\,10^{41}$ erg s$^{-1}$ (within a 
factor of 4).
This can be compared with $L_x$ expected from data by
Boer et al.\ (1992) who found optical evidence for an expanding supershell
with $R \sim 1.4$ kpc and d$R$/d$t \sim 250$ km s$^{-1}$ leading to an 
age of
$t=0.6 R/({\rm d}R/{\rm d}t) = 3.38\,10^6$ yrs . They measured a total 
H$\beta$ luminosity of $2.8\,10^{41}$ erg s$^{-1}$ 
yielding $L$(H$\alpha$)$ = 7.8\,10^{41}$ erg s$^{-1}$ in case B.
However, they estimate (on p.\ 74) that the outflowing shell contributes 
1\% of the H-recombination lines, but caution on p. 75 that this value 
might be
ten times higher due to incomplete coverage. 
Since the mechanical power should be hundred times 
that of H$\alpha$ (see Sect.\ 6.1.1) we obtain $L_{\rm mech} = 
7.8\,10^{41-42}$ erg s$^{-1}$
or a SN rate of 0.026 to 0.26 yr$^{-1}$.

Plugging the lower SN rate into eqs. (1) to (3) yields (for $n_{0.1}=1$)
$R = 1.3$ kpc, ${\rm d}R/{\rm d}t = 231$ km s$^{-1}$ and
$L_x = 8.6\,10^{39}$ erg s$^{-1}$. While $R$ and ${\rm d}R/{\rm d}t$ are
in good agreement with the measured data, $L_x$ is a factor 46 too small!
Increasing the uncertain parameter $n_{0.1}$ by a factor 10, would boost
the luminosity by a factor of only 3 and lower size and velocity of the 
shell
to 60\% the values they should have. The latter effect can be canceled
by taking the upper limit of $L_{\rm mech}$ by a factor of 10, leading 
to $L_x = 2.3\,10^{41}$ erg s$^{-1}$ which is consistent with the
measured $L_x$. 

The SN rate is comparable to that of M\,82 and NGC\,253
(Rieke et al.\ 1980; Kronberg et al. 1985), starburst galaxies with
an order of magnitude smaller $L_x$ from their centers
(Strickland et al. 1997). The difference is explained by the
hundred times smaller ambient density in these objects ($n_{0.1}=0.1$) 
than here assumed
for NGC\,2782 ($n_{0.1}=10$). If this latter conjecture will one day
turn out to be wrong then Seyfert (1943) might have been correct in
considering NGC\,2782 as a `Seyfert galaxy' of which we now see
some X-rays scattered in the outflow region.
However, presently the evidence for an AGN in NGC\,2782 does not appear
compelling.
\section{Concluding discussion}
We analyzed \ros data of the ultraluminous IR galaxy NGC\,6240 and 
of the starburst galaxy NGC\,2782. With our best fit of
the soft-X ray spectrum of NGC\,6240 we found a luminosity (0.1--2.4 keV)
of $5.2\,10^{42}$ erg s$^{-1}$ that, by at least an order of magnitude,
exceeds that of its ultraluminous
rival Arp\,220 (see Heckman et al.\ (1996): $4.3\,10^{40}$ to
$2.3\,10^{41}$ erg s$^{-1}$). The difference is partly due to a luminous
power-law component not present in Arp\,220. 
We interpret this power-law component as scattered
light from a hidden AGN with a luminosity that, depending on the poorly
known scattering geometry, provides a major fraction of the 
bolometric luminosity of NGC\,6240. About 20\% of the total $L_x$ in the 
\ros band
are due to thermal superwind-shocked shell material.

Lacking other components the scattered light of 
the hidden AGN is easiest to see in X rays. The scattered optical
continuum is hard to decompose from the strong
stellar background that was analyzed by Schmitt et al.\ (1996) and
Shier et al.\ (1996). Veilleux et al.\ (1995) noticed  
`featureless' continuum components in optical spectra of most ULIRGs. 
Polarization vectors presented in Radovich (1993) reveal a scattering 
geometry around the center
but more detailed work in this regard could be rewarding.

In Sect.\ 6.1.1 we saw that the hydrogen recombination-line 
spectrum is too strong to be completely attributed to shock heating. As 
shown by 
Schulz \& Fritsch (1994), attenuating the ionizing continuum of an AGN
by hidden ionized absorbing gaseous columns with, e.g. $\log N_{\rm H} 
\approx 20$, leads
automatically to a LINER like spectrum as observed in NGC\,6240.
This thin absorber is nearly transparent to X-rays.

Hence, the emerging picture of the core of NGC\,6240 is that of a buried
AGN component, which unlike a Seyfert with a fully developed
high-excitation spectrum, sees in no direction a completely clear sky.

NGC\,2782 has an order-of-magnitude smaller X-ray luminosity than
NGC\,6240 and can, in excellent agreement with optical data, be 
consistently 
explained by superbubble activity if the ambient density is relatively 
high
(average of 1 cm$^{-3}$ in the inner 2 kpc of the halo). This does not
completely exclude a hidden low-luminosity AGN, but so far, there is no
positive evidence for it.  

\begin{acknowledgements}
We thank an anonymous referee for pointing our attention to the
paper by Goldshmidt \& Rephaeli (1995).
H.S.\ gratefully acknowledges the hospitality of the Physics and
Astronomy Department of Dartmouth College
while finishing this work. T.W.B.\ acknowledges support from the
Alexander-von-Humboldt-Stiftung by a Feodor-Lynen Fellowship.
The \ros project is supported by the German Bundes\-mini\-ste\-rium
f\"ur Bildung, Wissenschaft, Forschung und Technologie (BMBF/DARA) and 
the Max-Planck-Society.
The optical images shown are based on photographic data of the 
National Geographic Society -- Palomar
Observatory Sky Survey (NGS-POSS) obtained using the Oschin Telescope on
Palomar Mountain. The NGS-POSS was funded by a grant from the National
Geographic Society to the California Institute of Technology.  The
plates were processed into the present compressed digital form with
their permission. The Digitized Sky Survey was produced at the Space
Telescope Science Institute under US Government grant NAG W-2166.
This research has made use of the NASA/IPAC extragalactic database (NED)
which is operated by the Jet Propulsion Laboratory, Caltech,
under contract with the National Aeronautics and Space
Administration.

\end{acknowledgements}

\end{document}